\documentstyle[aps,preprint]{revtex}

\tightenlines

\begin{document}

\title{\normalsize{\rm {\bf ANTISYMMETRIC TENSOR FIELDS, 
4-POTENTIALS AND INDEFINITE
METRICS}}\thanks{Presented at the {\it  5th Mexican School
``The Early Universe and Observational Cosmology."
Nov. 25-29, 2002, Playa del Carmen, M\'exico} and the {\it Jornadas de Investigaci\'on UAZ-2002, Zacatecas, Oct. 8-11, 2002.}}}

\author{{\bf Valeri V. Dvoeglazov}}

\address{{\rm Universidad de Zacatecas, Apartado Postal 636,
Suc. UAZ\\Zacatecas 98062, Zac., M\'exico\\
E-mail: valeri@ahobon.reduaz.mx\\
URL: http://ahobon.reduaz.mx/\~\,valeri/valeri.html}}

\date{November 16, 2002}


\maketitle

\begin{abstract}

\baselineskip14pt
We generalize the Stueckelberg formalism
in the $(1/2,1/2)$ representation of the Lorentz Group.
Some relations to other modern-physics models are found.
\end{abstract}

\newpage

\baselineskip15pt

\section{Outline.}

The plan of my talk is following:

\begin{itemize}

\item
Antecedents. Mapping between the Weinberg-Tucker-Hammer (WTH) formulation
and antisymmetric tensor (AST) fields of the 2nd rank. Modified Bargmann-Wigner (BW) formalism.  Pseudovector potential. Parity.

\item
Matrix form of the general equation in the $(1/2,1/2)$ representation.

\item
Lagrangian in the matrix form. Masses.

\item
Standard Basis and Helicity Basis.

\item
Dynamical invariants. Field operators. Propagators.

\item
Indefinite metric.

\end{itemize}

\section{Antecedents.}

Somebody may think that I am presenting well-known things. Therefore, I am going to give some overview of my previous works in order you to understand motivations better. In ref.~\cite{Hua}
I derived the Maxwell-like equations with the additional gradient of a scalar field $\chi$ from the first principles. Here they are:
\begin{mathletters}
\begin{eqnarray}
&&{\bbox \nabla}\times {\bf
E}=-\frac{1}{c}\frac{\partial {\bf B}}{\partial t} + {\bbox
\nabla} {\it Im} \chi \,, \label{1}\\
&&{\bbox \nabla }\times {\bf B}=\frac{1}{c}\frac{\partial {\bf
E}}{\partial t}  +{\bbox \nabla} {\it Re} \chi\,,\label{2}\\
&&{\bbox \nabla}\cdot {\bf E}=-{1\over c} {\partial \over \partial
t} {\it Re}\chi \,,\label{3}\\
&&{\bbox \nabla }\cdot {\bf B}= {1\over
c} {\partial \over \partial t} {\it Im} \chi \,.  \label{4}
\end{eqnarray}
\end{mathletters}
Of course, similar equations can be obtained 
in the massive case $m\neq 0$, i.e., within the Proca-like theory.
We should then consider
\begin{equation}
(E^2 -c^2 {\bf p}^2 - m^2 c^4 ) \Psi^{(3)} =0\, .\label{5}
\end{equation}
In the spin-1/2 case the equation (\ref{5}) can be written 
for the two-component spinor ($c=\hbar =1$)
\begin{equation}
(E I^{(2)} - {\bbox\sigma}\cdot {\bf p})
(E I^{(2)} + {\bbox\sigma}\cdot {\bf p})\Psi^{(2)} = m^2 \Psi^{(2)}\,,
\end{equation}
or, in the 4-component form\footnote{There exist various generalizations
of the Dirac formalism. For instance, the Barut generalization
is based on
\begin{equation}
[i\gamma_\mu \partial_\mu +  a (\partial_\mu \partial_\mu)/m - \ae ] \Psi =0\,,
\end{equation}
which can describe states of different masses. If one fixes the parameter $a$ by the requirement that the equation gives the state with the classical anomalous magnetic moment, then $m_2 =
m_1 (1+{3\over 2\alpha})$, i.e., it gives the muon mass. Of course, one can propose a generalized equation:
\begin{equation}
[i\gamma_\mu \partial_\mu +  a +b \Box + \gamma_5 (c+d\Box) ] \Psi =0
\,,
\end{equation}
$\Box= \partial_\mu \partial_\mu$; 
and, perhaps, even that of higher orders in derivatives.}
\begin{equation}
[i\gamma_\mu \partial_\mu +m_1 +m_2 \gamma^5 ] \Psi^{(4)} = 0\,.
\end{equation}
In the spin-1 case  we have
\begin{equation}
(E I^{(3)} - {\bf S}\cdot {\bf p})
(E I^{(3)} + {\bf S}\cdot {\bf p}){\bbox \Psi}^{(3)} 
- {\bf p} ({\bf p}\cdot {\bbox \Psi}^{(3)})= m^2 \Psi^{(3)}\,,
\end{equation}
that lead to (\ref{1}-\ref{4}), when $m=0$. We can continue writing down equations for higher spins in a similar fashion.

On this basis we are ready to generalize the BW formalism~\cite{bw-hs,Lurie}. Why is that convenient? In ref.~\cite{wig} I presented the mapping between the WTH equation~\cite{WTH} and the equations for AST fields. The equation for a 6-component field function is\footnote{ In order to have solutions satisfying the Einstein dispersion relations $E^2 -{\bf p}^2 =m^2$ we have to assume $B/(A+1)= 1$, or $B/(A-1)=1$.}
\begin{equation}
[\gamma_{\alpha\beta}p_\alpha p_\beta +A p_\alpha p_\alpha +Bm^2 ]
\Psi^{(6)} =0\,.
\end{equation}
Corresponding equations for the AST fields are:
\begin{mathletters}\begin{eqnarray}
&&\partial_\alpha\partial_\mu F_{\mu\beta}^{(I)}
-\partial_\beta\partial_\mu F_{\mu\alpha}^{(I)}
+ {A-1\over 2} \partial_\mu \partial_\mu F_{\alpha\beta}^{(I)}
-{B\over 2} m^2 F_{\alpha\beta}^{(I)} = 0\,,\label{wth1}\\
&&\partial_\alpha\partial_\mu F_{\mu\beta}^{(II)}
-\partial_\beta\partial_\mu F_{\mu\alpha}^{(II)}
- {A+1\over 2} \partial_\mu \partial_\mu F_{\alpha\beta}^{(II)}
+{B\over 2} m^2 F_{\alpha\beta}^{(II)} = 0\,,\label{wth2}
\end{eqnarray}\end{mathletters}
depending on the parity properties of $\Psi^{(6)}$ (the first case corresponds 
to the eigenvalue $P=-1$; the second one, to $P=+1$). 

We noted:

\begin{itemize}

\item
One can derive equations for the dual tensor $\tilde F_{\alpha\beta}$,
which are similar to (\ref{wth1},\ref{wth2}), refs.~[20a,7].

\item
In the Tucker-Hammer case ($A=1$, $B=2$), the first equation gives the Proca theory $\partial_\alpha \partial_\mu F_{\mu\beta} -
\partial_\beta \partial_\mu F_{\mu\alpha} = m^2 F_{\alpha\beta}$. In the second case one finds something different, $\partial_\alpha \partial_\mu F_{\mu\beta} -
\partial_\beta \partial_\mu F_{\mu\alpha} = (\partial_\mu \partial_\mu - m^2 ) F_{\alpha\beta}$

\item
If $\Psi^{(6)}$ has no definite parity, e.~g., $\Psi^{(6)} = 
\mbox{column} ({\bf E}+i{\bf B}\,\,\, {\bf B}+i{\bf E}\, )$, the equation
for the AST field will contain both the tensor and the dual tensor, e.~g.,
\begin{equation}
\partial_\alpha \partial_\mu F_{\mu\beta}
-\partial_\beta \partial_\mu F_{\mu\alpha}
={1\over 2} (\partial_\mu \partial_\mu) F_{\alpha\beta} +
[-{A\over 2} (\partial_\mu \partial_\mu) + {B\over 2} m^2] \tilde
F_{\alpha\beta}\,.\label{pv1}
\end{equation}

\item
Depending on the relation between $A$ and $B$ and on which
parity solution  do we consider, the WTH equations may describe different mass states. For instance, when $A=7$ and $B=8$ we have the second mass state $(m^{\prime})^2 = 4m^2/3$.

\end{itemize}

We tried to find relations between the generalized WTH theory
and other spin-1 formalisms.  Therefore, we were forced to modify the Bargmann-Wigner formalism~\cite{dv-ps,dv-cl}. For instance, we introduced the sign operator in the Dirac equations which are the input for the formalism for symmetric 2-rank spinor:
\begin{mathletters}
\begin{eqnarray}
\left [ i\gamma_\mu \partial_\mu + \epsilon_1 m_1 +\epsilon_2 m_2 \gamma_5
\right ]_{\alpha\beta} \Psi_{\beta\gamma} &=&0\,,\\
\left [ i\gamma_\mu
\partial_\mu + \epsilon_3 m_1 +\epsilon_4 m_2 \gamma_5 \right ]_{\gamma\beta}
\Psi_{\alpha\beta} &=&0\,,
\end{eqnarray}
\end{mathletters}
In general we have 16 possible combinations, but 4 of them give the same
sets of the Proca-like equations. We obtain~\cite{dv-cl}:\footnote{See the additional constraints in the cited paper.}
\begin{mathletters}
\begin{eqnarray} &&\partial_\mu A_\lambda - \partial_\lambda A_\mu + 2m_1 A_1 F_{\mu \lambda}
+im_2 A_2 \epsilon_{\alpha\beta\mu\lambda} F_{\alpha\beta} =0\,,\\
&&\partial_\lambda
F_{\mu \lambda} - {m_1\over 2} A_1 A_\mu -{m_2\over 2} B_2 \tilde
A_\mu=0\,,
\end{eqnarray} \end{mathletters}
with 
$A_1 = (\epsilon_1 +\epsilon_3) /2$,
$A_2 = (\epsilon_2 +\epsilon_4 )/ 2$,
$B_1 = (\epsilon_1 -\epsilon_3 )/ 2$,
and
$B_2 = (\epsilon_2 -\epsilon_4 )/ 2$.
So, we have the dual tensor and the pseudovector potential
in the Proca-like sets. The pseudovector potential is the same as that
which enters in the Duffin-Kemmer set for the spin 0. 

Moreover, it appears that the properties of the polarization
vectors with respect to parity operation depend on the choice of the spin basis.
For instance, in refs.~\cite{GR,dv-cl} the following momentum-space polarization vectors have been listed (in the pseudo-Euclidean metric):
\begin{mathletters}
\begin{eqnarray}
&&\epsilon _{\mu }({\bf p},\lambda =+1)={\frac{1}{\sqrt{2}}}{\frac{e^{i\phi }}{
p}}\pmatrix{ 0, {p_x p_z -ip_y p\over \sqrt{p_x^2 +p_y^2}}, {p_y p_z +ip_x
p\over \sqrt{p_x^2 +p_y^2}}, -\sqrt{p_x^2 +p_y^2}}\,, \\
&&\epsilon _{\mu }({\bf p},\lambda =-1)={\frac{1}{\sqrt{2}}}{\frac{e^{-i\phi }
}{p}}\pmatrix{ 0, {-p_x p_z -ip_y p\over \sqrt{p_x^2 +p_y^2}}, {-p_y p_z
+ip_x p\over \sqrt{p_x^2 +p_y^2}}, +\sqrt{p_x^2 +p_y^2}}\,, \\
&&\epsilon _{\mu }({\bf p},\lambda =0)={\frac{1}{m}}\pmatrix{ p, -{E \over p}
p_x, -{E \over p} p_y, -{E \over p} p_z }\,, \\
&&\epsilon _{\mu }({\bf p},\lambda =0_{t})={\frac{1}{m}}\pmatrix{E , -p_x,
-p_y, -p_z }\,.
\end{eqnarray}
\end{mathletters}
Berestetski\u{\i}, Lifshitz and Pitaevski\u{\i} claimed too~\cite{BLP} that
the helicity states cannot be the parity states. If one applies common-used
relations between fields and potentials it appears that the ${\bf E}$ and ${\bf B}$ fields have no ordinary properties with respect to space inversions:
\begin{mathletters}
\begin{eqnarray}
&&{\bf E}({\bf p},\lambda =+1)=-{\frac{iEp_{z}}{\sqrt{2}pp_{l}}}{\bf p}-{\frac{
E}{\sqrt{2}p_{l}}}\tilde{{\bf p}},\quad {\bf B}({\bf p},\lambda =+1)={\frac{
p_{z}}{\sqrt{2}p_{l}}}{\bf p}-{\frac{ip}{\sqrt{2}p_{l}}}\tilde{{\bf p}}, \nonumber\\
&&\\
&&{\bf E}({\bf p},\lambda =-1)=+{\frac{iEp_{z}}{\sqrt{2}pp_{r}}}{\bf p}-{\frac{
E}{\sqrt{2}p_{r}}}\tilde{{\bf p}}^{\ast },\quad {\bf B}({\bf p},\lambda
=-1)={\frac{p_{z}}{\sqrt{2}p_{r}}}{\bf p}+{\frac{ip}{\sqrt{2}p_{r}}}\tilde{
{\bf p}}^{\ast }, \nonumber\\
&&\\
&&{\bf E}({\bf p},\lambda =0)={\frac{im}{p}}{\bf p},\quad {\bf B}({\bf p}
,\lambda =0)=0,
\end{eqnarray}
\end{mathletters}
with $\tilde {\bf p}=\pmatrix{p_y\cr -p_x\cr -ip\cr}$.

Thus, the conclusions of our previous works are:
\begin{itemize}

\item
There exists the mapping between the WTH formalism for $S=1$ and the AST fields of four kinds (provided that the solutions of the WTH equations are of the definite
parity).

\item
Their massless limits contain additional solutions comparing with the Maxwell equations. This was related to the possible theoretical existence of the Ogievetski\u{\i}-Polubarinov-Kalb-Ramond notoph~\cite{Og}.

\item
In some particular cases ($A=0, B=1$) massive solutions of different parities are naturally divided into the classes of causal and tachyonic solutions.

\item
If we want to take into account the solutions of the WTH equations of
different parity properties, this induces us to generalize the BW, Proca and Duffin-Kemmer formalisms.

\item
In the $(1/2,0)\oplus (0,1/2)$, $(1,0)\oplus (0,1)$ etc. representations
it is possible to introduce the parity-violating frameworks. The corresponding solutions are the mixture of various polarization states.

\item
The addition of the Klein-Gordon equation to the $(S,0)\oplus (0,S)$ equations may change the theoretical content even on the free level. For instance, the higher-spin equations may actually describe various spin and mass states.

\item
There also exist the mappings between the WTH solutions of undefined parity
and the AST fields, which contain both tensor and dual tensor. They are eight.

\item
The 4-potentials and electromagnetic fields~\cite{GR,dv-cl} in the helicity
basis have different parity properties comparing with the standard basis of the polarization vectors.

\item
In the previous talk~\cite{dv-pl} we presented a theory in the $(1/2,0)\oplus (0,1/2)$ representation in the helicity basis. Under space inversion operation,
different helicity states transform each other, $Pu_h (-{\bf p}) = -i u_{-h} ({\bf p})$, $Pv_h (-{\bf p}) = +i v_{-h} ({\bf p})$.

\end{itemize}

I hope, this is enough for the antecedents. Everybody has already understood the importance of 
$\tilde A_\mu \sim \partial_\mu \chi$ term in the electrodynamics and in the Proca theory.

\section{The theory of 4-vector field.}

First of all, we show that the equation for the 4-vector field can be presented
in a matrix form. Recently, S. I. Kruglov proposed~\cite{krug1,krug2}\footnote{I acknowledge the discussion of physical significance of the gauge 
with M. Kirchbach in 1998. See also: R. A. Berg, Nuovo Cim. A \, {\bf XLII}, 148 (1966) and D. V. Ahluwalia and M. Kirchbach, Mod. Phys. Lett. A{\bf 16}, 1377 (2001).} a general form of the Lagrangian for  4-potential field $B_\mu$, which also contains the spin-0 state. 
Initially, we have (provided that derivatives commute)
\begin{equation}
\alpha \partial_\mu \partial_\nu B_\nu +\beta \partial_\nu^2 B_\mu +\gamma m^2 B_\mu =0\, .\label{eq-pot}
\end{equation}
When $\partial_\nu B_\nu =0$ (the Lorentz gauge) we obtain spin-1 states only.
However, if it is not equal to zero we have a scalar field and a pseudovector potential. We can also check this by consideration of the dispersion relations of (\ref{eq-pot}). One obtains 4+4 states (two of them may differ in mass
from others).

Next, one can fix one of the constants $\alpha,\beta,\gamma$ 
without loosing any physical content. For instance, when $\alpha=-2$
and taking into account that the action of the symmetrized combination
of Kronecker's $\delta$'s \, is
\begin{equation}
(\delta_{\mu\nu}\delta_{\alpha\beta} - \delta_{\mu\alpha} \delta_{\nu\beta}
- \delta_{\mu\beta} \delta_{\nu\alpha}) \partial_\alpha \partial_\beta
B_\nu = \partial_\alpha^2 B_\mu - 2\partial_\mu \partial_\nu B_\nu\,,
\end{equation}
one gets the equation
\begin{equation}
\left [ \delta_{\mu\nu} \delta_{\alpha\beta} - \delta_{\mu\alpha}\delta_{\nu\beta} - \delta_{\mu\beta} \delta_{\nu\alpha}\right ] \partial_\alpha \partial_\beta B_\nu + A \partial_\alpha^2 \delta_{\mu\nu} B_\nu - Bm^2 \delta_{\mu\nu} B_\nu =0\,,\label{eq1-m}
\end{equation} 
where  $\beta= A+1$ and $\gamma=-B$. In the matrix form the equation (\ref{eq1-m}) reads:
\begin{equation}
\left [ \gamma_{\alpha\beta} \partial_\alpha \partial_\beta +A \partial_\alpha^2 - Bm^2 \right ]_{\mu\nu} B_\nu = 0\,,
\end{equation}
with
\begin{equation}
[\gamma_{\alpha\beta}]_{\mu\nu} = \delta_{\mu\nu}\delta_{\alpha\beta}
-\delta_{\mu\alpha}\delta_{\nu\beta} - \delta_{\mu\beta} \delta_{\nu\alpha}\,.
\end{equation}
Their explicit forms are the following ones:
\begin{mathletters}\begin{eqnarray}
&&\gamma_{44}=\pmatrix{1&0&0&0\cr
0&1&0&0\cr
0&0&1&0\cr
0&0&0&-1\cr}\,,\quad
\gamma_{14}=\gamma_{41}=\pmatrix{0&0&0&-1\cr
0&0&0&0\cr
0&0&0&0\cr
-1&0&0&0\cr}\,,\\
&&\gamma_{24}=\gamma_{42}=\pmatrix{0&0&0&0\cr
0&0&0&-1\cr
0&0&0&0\cr
0&-1&0&0\cr}\,,\quad
\gamma_{34}=\gamma_{43}=\pmatrix{0&0&0&0\cr
0&0&0&0\cr
0&0&0&-1\cr
0&0&-1&0\cr}\,,\\
&&\gamma_{11}=\pmatrix{-1&0&0&0\cr
0&1&0&0\cr
0&0&1&0\cr
0&0&0&1\cr}\,,\quad
\gamma_{22}=\pmatrix{1&0&0&0\cr
0&-1&0&0\cr
0&0&1&0\cr
0&0&0&1\cr}\,,\\
&&\gamma_{33}=\pmatrix{1&0&0&0\cr
0&1&0&0\cr
0&0&-1&0\cr
0&0&0&1\cr}\,,\quad
\gamma_{12}=\gamma_{21}=\pmatrix{0&-1&0&0\cr
-1&0&0&0\cr
0&0&0&0\cr
0&0&0&0\cr}\,,\\
&&\gamma_{31}=\gamma_{13}=\pmatrix{0&0&-1&0\cr
0&0&0&0\cr
-1&0&0&0\cr
0&0&0&0\cr}\,,\quad
\gamma_{23}=\gamma_{32}=\pmatrix{0&0&0&0\cr
0&0&-1&0\cr
0&-1&0&0\cr
0&0&0&0\cr}\,.
\end{eqnarray}
\end{mathletters}
They are the analogs of the Barut-Muzinich-Williams (BMW) $\gamma$-matrices
for bivector fields. However, $\sum_{\alpha}^{} [\gamma_{\alpha\alpha}]_{\mu\nu} = 
2\delta_{\mu\nu}$. It is easy to prove by the textbook method~\cite{Itzyk} that $\gamma_{44}$
can serve as the parity matrix.

One can also define the analogs of the BMW $\gamma_{5,\alpha\beta}$ matrices 
\begin{equation}
\gamma_{5,\alpha\beta} = {i\over 6} [\gamma_{\alpha\kappa}, \gamma_{\beta\kappa} ]_{-, \mu\nu} = i [\delta_{\alpha\mu} \delta_{\beta\nu} - \delta_{\alpha\nu}\delta_{\beta\mu} ]\,.
\end{equation}
As opposed to $\gamma_{\alpha\beta}$ matrices they are totally anti-symmetric.
The explicit forms of the anti-symmetric $\gamma_{5,\alpha\beta}$ are\footnote{They are related to  boost and rotation generators of this representation.}
\begin{mathletters}
\begin{eqnarray}
&&\gamma_{5,41} = - \gamma_{5,14} = i \pmatrix{0&0&0&-1\cr
0&0&0&0\cr
0&0&0&0\cr
1&0&0&0\cr}\,,\quad
\gamma_{5,42} = - \gamma_{5,24} = i \pmatrix{0&0&0&0\cr
0&0&0&-1\cr
0&0&0&0\cr
0&1&0&0\cr}\,,\nonumber\\
&&\\
&&\nonumber\\
&&\gamma_{5,43} = - \gamma_{5,34} = i \pmatrix{0&0&0&0\cr
0&0&0&0\cr
0&0&0&-1\cr
0&0&1&0\cr}\,,\quad
\gamma_{5,12} = - \gamma_{5,21} = i \pmatrix{0&1&0&0\cr
-1&0&0&0\cr
0&0&0&0\cr
0&0&0&0\cr}\,,\nonumber\\
&&\\
&&\gamma_{5,31} = - \gamma_{5,13} = i \pmatrix{0&0&-1&0\cr
0&0&0&0\cr
1&0&0&0\cr
0&0&0&0\cr}\,,\quad
\gamma_{5,23} = - \gamma_{5,32} = i \pmatrix{0&0&0&0\cr
0&0&1&0\cr
0&-1&0&0\cr
0&0&0&0\cr}\,.\nonumber\\
&&
\end{eqnarray}
\end{mathletters}
$\gamma$-matrices are pure real; $\gamma_5$-matrices are pure imaginary.
In the $(1/2,1/2)$ representation, we need 16 matrices to form the complete set (as opposed to the bi-vector representation, when we have to define also
$\gamma_{6,\alpha\beta,\mu\nu}$).
Please note that in the pseudo-Euclidean metric the symmetry properties of the $\gamma$'s and $\gamma_5$'s are {\it not} the same  (comparing with our consideration in the Euclidean metric) in such a representation.

\medskip

{\it Lagrangian and the equations of motion.}
Let us try
\begin{equation}
{\cal L} = (\partial_\alpha B_\mu^\ast) [\gamma_{\alpha\beta} ]_{\mu\nu} (\partial_\beta B_\nu)
+ A (\partial_\alpha B_\mu^\ast) (\partial_\alpha B_\mu) + Bm^2
B_\mu^\ast B_\mu\,.
\end{equation}
On using the Lagrange-Euler equation
\begin{equation}
\frac{\partial {\cal L}}{\partial B_\mu^\ast}
- \partial_\nu (\frac{\partial {\cal L}}{\partial (\partial_\nu B_\mu^\ast)})
=0\,,
\end{equation}
or
\begin{equation}
\frac{\partial {\cal L}}{\partial B_\mu}
- \partial_\nu (\frac{\partial {\cal L}}{\partial (\partial_\nu B_\mu)})
=0\,,
\end{equation}
we have
\begin{equation}
[\gamma_{\nu\beta}]_{\kappa\tau} \partial_\nu \partial_\beta B_\tau + A \partial_\nu^2 B_\kappa - Bm^2  B_\kappa =0\,,\label{equat}
\end{equation}
or
\begin{equation}
[\gamma_{\beta\nu}]_{\kappa\tau} \partial_\beta \partial_\nu B_\tau^\ast + A \partial_\nu^2 B_\kappa^\ast - Bm^2  B_\kappa^\ast =0\,.\label{eq-f}
\end{equation}
Thus, they may be presented in the form of (15). The Lagrangian is correct. 

\medskip

{\it Masses.} We are convinced that in the case of spin 0, we have $B_\mu \rightarrow
\partial_\mu \chi$; in the case of spin 1 we have $\partial_\mu B_\mu =0$.

So,
\begin{enumerate}

\item

\begin{equation}
(\delta_{\mu\nu}\delta_{\alpha\beta}
-\delta_{\mu\alpha}\delta_{\nu\beta} - \delta_{\mu\beta} \delta_{\nu\alpha})
\partial_\alpha \partial_\beta \partial_\nu \chi = - \partial^2 \partial_\mu \chi\,.
\end{equation}
Hence, from (\ref{equat}) we have
\begin{equation}
[(A-1) \partial^2_\nu - Bm^2 ] \partial_\mu \chi=0\,.
\end{equation}
If $A-1=B$ we have the spin-0 particles with masses $\pm m$ with the correct relativistic dispersion.

\smallskip

\item
In another case
\begin{equation}
[\delta_{\mu\nu}\delta_{\alpha\beta}
-\delta_{\mu\alpha}\delta_{\nu\beta} - \delta_{\mu\beta} \delta_{\nu\alpha}]
\partial_\alpha \partial_\beta B_\nu  =  \partial^2 B_\mu \,.
\end{equation}
Hence,
\begin{equation}
[(A+1) \partial^2_\nu  - Bm^2] B_\mu =0\,.
\end{equation}
If $A+1 =B$ we have the spin-1 particles with masses $\pm m$ with the correct relativistic dispersion.

\end{enumerate}

The equation (\ref{equat}) can be transformed in  two equations:
\begin{mathletters}
\begin{eqnarray}
\left [\gamma_{\alpha\beta} \partial_\alpha \partial_\beta + (B+1)\partial_\alpha^2 - Bm^2 \right ]_{\mu\nu} B_\nu &=&0\,, \quad \mbox{spin 0 with masses}\, \pm m\,,\\
\left [\gamma_{\alpha\beta} \partial_\alpha \partial_\beta + (B-1)\partial_\alpha^2  - Bm^2 \right ]_{\mu\nu}  B_\nu &=&0\,, \quad \mbox{spin 1 with masses}\, \pm m\,.
\end{eqnarray}
\end{mathletters}
The first one has the solution with spin 0 and masses $\pm m$. However, it has also the {\it spin-1} solution with the {\it different masses}, $[\partial_\nu^2 +(B+1)\partial^2_\nu - Bm^2 ] B_\mu =0$:
\begin{equation}
\tilde m = \pm \sqrt{{B\over B+2}} m\,.
\end{equation}
The second one has the solution with spin 1 and masses $\pm m$. But, it also has 
the {\it spin-0} solution with the {\it different masses}, $[ -\partial_\nu^2 + (B-1) \partial^2_\nu - Bm^2 ] \partial_\mu \chi =0$:
\begin{equation}
\tilde m = \pm \sqrt{{B\over B-2}}m\,.
\end{equation}
One can come to the same conclusion by checking the dispersion relations 
from $\mbox{Det} [\gamma_{\alpha\beta} p_\alpha p_\beta - Ap_\alpha p_\alpha +Bm^2] = 0$\,. When $\tilde m^2 = {4\over 3} m^2$, we have $B=-8, A=-7$, that is compatible with our consideration of bi-vector fields~\cite{wig}.

One can form the Lagrangian with the particles of spines 1, masses $\pm m$, the particle with the mass $\sqrt{{4\over 3}} m$, spin 1, for which the particle is equal to the antiparticle, by choosing the appropriate creation/annihilation operators; and the particles with spines 0 with masses $\pm m$ and 
$\pm \sqrt{{4\over 5}} m$ (some of them may be neutral).

\medskip

{\it The Standard Basis}~\cite{Novozh,Weinb,Dv-book}.
The polarization vectors of the standard basis are defined:
\begin{mathletters}
\begin{eqnarray}
&&\epsilon_\mu
({\bf 0}, +1)= -{1\over \sqrt{2}}\pmatrix{1\cr i\cr 0 \cr 0\cr}\,,\quad  
\epsilon_\mu ({\bf 0}, -1)= +{1\over
\sqrt{2}}\pmatrix{1\cr -i\cr 0\cr 0\cr}\,,\\
&&\epsilon_\mu ({\bf
0}, 0) = \pmatrix{0\cr 0\cr 1\cr 0\cr}\,,\quad
\epsilon_\mu ({\bf
0}, 0_t) = \pmatrix{0\cr 0\cr 0\cr i\cr}\,.
\end{eqnarray}
\end{mathletters}
The Lorentz transformations are:
\begin{mathletters}
\begin{eqnarray} &&
\epsilon_\mu ({\bf p}, \sigma) =
L_{\mu\nu} ({\bf p}) \epsilon_\nu ({\bf 0},\sigma)\,,\\ 
&& L_{44} ({\bf p}) = \gamma\, ,\, L_{i4} ({\bf p}) = -
L_{4i} ({\bf p}) = i\widehat p_i \sqrt{\gamma^2 -1}\, ,\,
L_{ik} ({\bf p}) = \delta_{ik} + (\gamma -1) \widehat p_i \widehat
p_k \,. \end{eqnarray} \end{mathletters}
Hence, for the particles of the mass $m$ we have:
\begin{mathletters} \begin{eqnarray} 
u^\mu
({\bf p}, +1)&=& -{N\over \sqrt{2}m}\pmatrix{m+ {p_1 p_r \over
E_p+m}\cr im +{p_2 p_r \over E_p+m}\cr {p_3 p_r \over
E_p+m}\cr -ip_r\cr}\,,\quad u^\mu ({\bf p}, -1)= {N\over
\sqrt{2}m}\pmatrix{m+ {p_1 p_l \over E_p+m}\cr -im +{p_2 p_l \over
E_p+m}\cr {p_3 p_l \over E_p+m}\cr -ip_l\cr}\,,\nonumber\\
&&\label{vp12}\\ u^\mu ({\bf
p}, 0) &=& {N\over m}\pmatrix{{p_1 p_3 \over E_p+m}\cr {p_2 p_3
\over E_p+m}\cr m + {p_3^2 \over E_p+m}\cr -ip_3\cr}\,,\quad
u^\mu ({\bf p}, 0_t) = {N \over m} \pmatrix{-p_1
\cr -p_2\cr -p_3\cr iE_p\cr}\,.
\end{eqnarray}
\end{mathletters}
The Euclidean metric was again used; $N$ is the normalization constant. They are the eigenvectors of the parity operator:
\begin{equation}
Pu_\mu (-{\bf p}, \sigma) = + u_\mu ({\bf p}, \sigma)\,,\quad
Pu_\mu (-{\bf p}, 0_t) = -u_\mu ({\bf p}, 0_t)\,.
\end{equation}

\medskip

{\it The Helicity Basis.}~\cite{GR,Car}
The helicity operator is:
\begin{equation}
{({\bf J}\cdot {\bf p})\over p} = {1\over p} \pmatrix{
0&-ip_z&ip_y&0\cr
ip_z&0&-ip_x&0\cr
-ip_y&ip_x&0&0\cr
0&0&0&0\cr}\,,\,\,
{({\bf J}\cdot {\bf p})\over p} \epsilon^\mu_{\pm 1} = \pm \epsilon^\mu_{\pm 1}\,,\,\,{({\bf J}\cdot {\bf p})\over p} \epsilon^\mu_{0,0_t} = 0\,.
\end{equation}
The eigenvectors are:
\begin{mathletters}
\begin{eqnarray}
&&\epsilon^\mu_{+1}= {1\over \sqrt{2}} {e^{i\alpha}\over p} \pmatrix{
{-p_x p_z +ip_y p\over \sqrt{p_x^2 +p_y^2}}\cr {-p_y p_z -ip_x
p\over \sqrt{p_x^2 +p_y^2}}\cr \sqrt{p_x^2 +p_y^2}\cr 0\cr}\,,\quad
\epsilon^{\mu }_{-1}={1\over \sqrt{2}}{e^{i\beta}\over p}
\pmatrix{{p_x p_z +ip_y p\over \sqrt{p_x^2 +p_y^2}}\cr {p_y p_z
-ip_x p\over \sqrt{p_x^2 +p_y^2}}\cr -\sqrt{p_x^2 +p_y^2}\cr 0\cr}\,, \\
&&\epsilon^{\mu }_0={\frac{1}{m}}\pmatrix{{E \over p}
p_x \cr {E \over p} p_y \cr{E \over p} p_z\cr ip }\,,\quad
\epsilon^{\mu }_{0_{t}}={\frac{1}{m}}\pmatrix{p_x\cr
p_y\cr p_z\cr iE_p\cr}\,.
\end{eqnarray}
\end{mathletters}
The eigenvectors $\epsilon^\mu_{\pm  1}$ are not the eigenvectors
of the parity operator ($\gamma_{44}$) of this representation. However, $\epsilon^\mu_{1,0}$,
$\epsilon^\mu_{0,0_t}$
are. Surprisingly, the latter have no well-defined massless limit.\footnote{In order to get the well-known massless limit one should use the basis of the light-front form reprersentation, cf.~\cite{Ahl-lf}.}

\medskip

{\it Energy-momentum tensor.}
According to definitions~\cite{Lurie} it is defined as
\begin{mathletters}
\begin{eqnarray}
&&T_{\mu\nu} = - \sum_{\alpha}^{} \left [ {\partial {\cal L} \over \partial (\partial_\mu B_\alpha)} \partial_\nu B_\alpha
+\partial_\nu B_\alpha^\ast {\partial {\cal L} \over \partial (\partial_\mu B_\alpha^\ast)}\right ]
+{\cal L} \delta_{\mu\nu}\,,\\
&& P_\mu = -i \int T_{4\mu} d^3 {\bf x}\,.
\end{eqnarray}
\end{mathletters}
Hence,
\begin{eqnarray}
&&T_{\mu\nu} =  -(\partial_\kappa B_\tau^\ast) [\gamma_{\kappa\mu}]_{\tau\alpha} (\partial_\nu B_\alpha) - (\partial_\nu B_\alpha^\ast) [\gamma_{\mu\kappa}]_{\alpha\tau} (\partial_\kappa B_\tau)-\nonumber\\
&-& A [(\partial_\mu B_\alpha^\ast) (\partial_\nu B_\alpha) + (\partial_\nu B_\alpha^\ast)  (\partial_\mu B_\alpha)] + {\cal L}\delta_{\mu\nu} =\nonumber\\
&=& - (A+1) [(\partial_\mu B_\alpha^\ast) (\partial_\nu B_\alpha) + (\partial_\nu B_\alpha^\ast)  (\partial_\mu B_\alpha)] +
\left [ (\partial_\alpha B_\mu^\ast) (\partial_\nu B_\alpha) + \right . \nonumber\\
&+&\left . (\partial_\nu B_\alpha^\ast)  (\partial_\alpha B_\mu) \right ] +
[(\partial_\alpha B_\alpha^\ast) (\partial_\nu B_\mu) + (\partial_\nu B_\mu^\ast)  (\partial_\alpha B_\alpha)] + {\cal L} \delta_{\mu\nu}\,.
\end{eqnarray}
Remember that after substitutions of  the explicite forms $\gamma$'s, the Lagrangian is
\begin{equation}
{\cal L} = (A+1) (\partial_\alpha B_\mu^\ast) (\partial_\alpha B_\mu ) - (\partial_\nu B_\mu^\ast)  (\partial_\mu B_\nu)- (\partial_\mu B_\mu^\ast) (\partial_\nu B_\nu) + Bm^2   B_\mu^\ast B_\mu\,,
\end{equation}
and the third term cannot be removed by the standard substitution ${\cal L} \rightarrow {\cal L}^\prime +\partial_\mu \Gamma_\mu$\,,$\Gamma_\mu = B_\nu^\ast \partial_\nu B_\mu - B_\mu^\ast \partial_\nu B_\nu$ 
to get the textbook Lagrangian ${\cal L}^\prime = (\partial_\alpha B_\mu^\ast) (\partial_\alpha B_\mu ) +m^2 B_\mu^\ast B_\mu$\,.

\medskip

{\it The current vector} is defined
\begin{mathletters}
\begin{eqnarray}
&&J_{\mu} = -i \sum_{\alpha}^{} [{\partial {\cal L} \over \partial (\partial_\mu B_\alpha)} 
B_\alpha
- B_\alpha^\ast {\partial {\cal L} \over \partial (\partial_\mu B_\alpha^\ast)} ]\,,\\
&& Q = -i \int J_{4} d^3 {\bf x}\,.
\end{eqnarray}
\end{mathletters}
Hence,
\begin{eqnarray}
&&J_{\lambda} =  -i \left \{ (\partial_\alpha B_\mu^\ast) [\gamma_{\alpha\lambda}]_{\mu\kappa}  B_\kappa - B_\kappa^\ast [\gamma_{\lambda\alpha}]_{\kappa\mu} (\partial_\alpha B_\mu) +
A (\partial_\lambda B_\kappa^\ast) B_\kappa - A B_\kappa^\ast (\partial_\lambda B_\kappa) \right \} \nonumber\\
&=& - i \left \{ (A+1) [(\partial_\lambda B_\kappa^\ast) B_\kappa
 -  B_\kappa^\ast (\partial_\lambda B_\kappa) ] +  [ B_\kappa^\ast (\partial_\kappa B_\lambda) -
(\partial_\kappa B_\lambda^\ast) B_\kappa ] + \right . \nonumber\\
&+&  \left . [B_\lambda^\ast (\partial_\kappa B_\kappa) - (\partial_\kappa B_\kappa^\ast) B_\lambda ] \right \} \,.
\end{eqnarray}
Again, the second term and the last term cannot be removed at the same time by adding the total derivative to the Lagrangian. These terms correspond to the contribution of the scalar (spin-0) portion.

\medskip

{\it Angular momentum.}
Finally,
\begin{mathletters}
\begin{eqnarray}
&&{\cal M}_{\mu\alpha,\lambda} = x_\mu T_{\{\alpha\lambda\}} - x_\alpha T_{\{\mu\lambda\}} + {\cal S}_{\mu\alpha,\lambda} =\nonumber\\
&=& x_\mu T_{\{\alpha\lambda\}} - x_\alpha T_{\{\mu\lambda\}} -  i \left \{\sum_{\kappa\tau}^{}
{\partial {\cal L} \over \partial (\partial_\lambda B_\kappa)} {\cal T}_{\mu\alpha,\kappa\tau} B_\tau+ B_\tau^\ast {\cal T}_{\mu\alpha,\kappa\tau} {\partial {\cal L} \over 
\partial (\partial_\lambda B_\kappa^\ast)}\right \}\,,\\
&& {\cal M}_{\mu\nu} = -i \int {\cal M}_{\mu\nu,4} d^3 {\bf x}\,,
\end{eqnarray}
\end{mathletters}
where ${\cal T}_{\mu\alpha,\kappa\tau} \sim [\gamma_{5,\mu\alpha}]_{\kappa\tau}$\,.

\medskip

{\it The field operator.} Various-type field operators are possible in this representation. Let us remind the textbook procedure to get them.
During the calculations below we have to present $1=\theta (k_0) +\theta (-k_0)$
in order to get positive- and negative-frequency parts. However, one should be warned that in the point $k_0=0$ this presentation is ill-defined.
\begin{eqnarray}
&&A_\mu (x) = {1\over (2\pi)^3} \int d^4 k \,\delta (k^2 -m^2) e^{+ik\cdot x}
A_\mu (k) =\nonumber\\
&=& {1\over (2\pi)^3} \sum_{\lambda}^{}\int d^4 k \delta (k_0^2 -E_k^2) e^{+ik\cdot x}
\epsilon_\mu (k,\lambda) a_\lambda (k) =\nonumber\\
&=&{1\over (2\pi)^3} \int {d^4 k \over 2E_k} [\delta (k_0 -E_k) +\delta (k_0 +E_k) ] 
[\theta (k_0) +\theta (-k_0) ]e^{+ik\cdot x}
A_\mu (k) =\nonumber\\
&=&{1\over (2\pi)^3} \int {d^4 k \over 2E_k} [\delta (k_0 -E_k) +\delta (k_0 +E_k) ] \left
[\theta (k_0) A_\mu (k) e^{+ik\cdot x}  + \right.\nonumber\\
&+&\left.\theta (k_0) A_\mu (-k) e^{-ik\cdot x} \right ]
={1\over (2\pi)^3} \int {d^3 {\bf k} \over 2E_k} \theta(k_0)  
[A_\mu (k) e^{+ik\cdot x}  + A_\mu (-k) e^{-ik\cdot x} ]
=\nonumber\\
&=&{1\over (2\pi)^3} \sum_{\lambda}^{}\int {d^3 {\bf k} \over 2E_k}   
[\epsilon_\mu (k,\lambda) a_\lambda (k) e^{+ik\cdot x}  + \epsilon_\mu (-k,\lambda) 
a_\lambda (-k) e^{-ik\cdot x} ]\,.
\end{eqnarray}
Moreover, we should transform the second part to $\epsilon_\mu^\ast (k,\lambda) b_\lambda^\dagger (k)$ as usual. In such a way we obtain the charge-conjugate states. Of course, one can try to get $P$-conjugates or $CP$-conjugate states too. One should proceed in a similar way as in the Appendix.
We set
\begin{equation}
\sum_{\lambda}^{} \epsilon_\mu (-k,\lambda) a_\lambda (-k) = 
\sum_{\lambda}^{} \epsilon_\mu^\ast (k,\lambda) b_\lambda^\dagger (k)\,,
\label{expan}
\end{equation}
multiply both parts by $\epsilon_\nu [\gamma_{44}]_{\nu\mu}$, and use the normalization conditions for polarization vectors.

In the $({1\over 2}, {1\over 2})$ representation we can also expand
(apart the equation (\ref{expan})) in the different way:
\begin{equation}
\sum_{\lambda}^{} \epsilon_\mu (-k, \lambda) a_\lambda (-k) =
\sum_{\lambda}^{} \epsilon_\mu (k, \lambda) a_\lambda (k)\,.
\end{equation}
From the first definition we obtain (the signs $\mp$
depends on the value of $\sigma$):
\begin{equation}
b_\sigma^\dagger (k) = \mp \sum_{\mu\nu\lambda}^{} \epsilon_\nu (k,\sigma) 
[\gamma_{44}]_{\nu\mu} \epsilon_\mu (-k,\lambda) a_\lambda (-k)\,,
\end{equation}
or
\begin{eqnarray}
b_\sigma^\dagger (k) = {E_k^2 \over m^2} \pmatrix{1+{{\bf k}^2\over E_k^2}&\sqrt{2}
{k_r \over E_k}&-\sqrt{2} {k_l \over E_k}& -{2k_3 \over E_k}\cr
-\sqrt{2} {k_r \over E_k}&-{k_r^2 \over {\bf k}^2}& -{m^2k_3^2\over E_k^2 {\bf k}^2}
+{k_r k_l \over E_k^2} & {\sqrt{2} k_3 k_r \over {\bf k}^2}\cr
\sqrt{2} {k_l \over E_k}&-{m^2 k_3^2 \over E_k^2 {\bf k}^2} + {k_r k_l \over E_k^2}& -{k_l^2\over {\bf k}^2} & -{\sqrt{2} k_3 k_l \over {\bf k}^2}\cr
{2k_3 \over E_k}&{\sqrt{2}k_3 k_r \over {\bf k}^2}& -{\sqrt{2} k_3 k_l\over {\bf k}^2} & {m^2 \over E_k^2} -{2 k_3 \over {\bf k}^2}\cr}
\pmatrix{a_{00} (-k)\cr a_{11} (-k)\cr
a_{1-1} (-k)\cr a_{10} (-k)\cr}\,.\nonumber
\end{eqnarray}
\begin{equation}
.
\end{equation}
From the second definition $\Lambda^2_{\sigma\lambda} = \mp \sum_{\nu\mu}^{} \epsilon^{\ast}_\nu (k, \sigma) [\gamma_{44}]_{\nu\mu}
\epsilon_\mu (-k, \lambda)$ we have
\begin{eqnarray}
a_\sigma (k) =  \pmatrix{-1&0&0&0\cr
0&{k_3^2 \over {\bf k}^2}& {k_l^2\over {\bf k}^2} & {\sqrt{2} k_3 k_l \over {\bf k}^2}\cr
0&{k_r^2 \over {\bf k}^2}& {k_3^2\over {\bf k}^2} & -{\sqrt{2} k_3 k_r \over {\bf k}^2}\cr
0&{\sqrt{2}k_3 k_r \over {\bf k}^2}& -{\sqrt{2} k_3 k_l\over {\bf k}^2} & 1-{2 k_3^2 \over {\bf k}^2}\cr}\pmatrix{a_{00} (-k)\cr a_{11} (-k)\cr
a_{1-1} (-k)\cr a_{10} (-k)\cr}\,.
\end{eqnarray}
It is the strange case: the field operator will only destroy particles. Possibly, we should think about modifications of the Fock space in this case, or introduce several field operators for the $({1\over 2}, {1\over 2})$ representation.

\medskip

{\it Propagators.} From ref.~\cite{Itzyk} it is known for the real vector field:
\begin{equation}
<0\vert T(B_\mu (x) B_\nu (y)\vert 0> = -i \int {d^4 k \over (2\pi)^4} e^{ik (x-y)} 
(\frac{\delta_{\mu\nu} +k_\mu k_\nu/\mu^2}{k^2 +\mu^2 +i\epsilon} - \frac{k_\mu k_\nu/\mu^2}{k^2 +m^2 +i\epsilon})\,.
\end{equation}
If $\mu=m$ (this depends on relations between $A$ and  $B$) we have the cancellation of divergent parts. Thus, we can overcome the well-known  difficulty of the Proca theory with the massless limit. 

If $\mu\neq m$ we can still have a {\it causal} theory, but in this case we need more than one equation, and should apply the method proposed 
in ref.~\cite{dv-hpa}.\footnote{In that case we applied  for the bi-vector fields
\begin{eqnarray}
&&\hspace*{-1cm}\left [ \gamma_{\mu\nu} \partial_\mu \partial_\nu -m^2 \right ]
 \int  \frac{d^3 {\bf p}}{(2\pi)^3 8im^2 E_p}
\left [ \theta (t_2 -t_1) u^1_{\sigma} ({\bf p}) \otimes \overline
u^1_{\sigma} ({\bf p}) e^{ip\cdot x}+\right .\nonumber\\
&&\left.  \qquad\qquad+\theta (t_1 -t_2) v^1_{\sigma} ({\bf p})
\otimes \overline  v^1_{\sigma} ({\bf p}) e^{-ipx} \right  ] +\\
&+& \left [ \gamma_{\mu\nu} \partial_\mu \partial_\nu + m^2 \right  ]  \int
\frac{d^3 p}{(2\pi)^3 8im^2 E_p}
\left [ \theta (t_2 -t_1) u^2_{\sigma} ({\bf p}) \otimes \overline
u^2_{\sigma} ({\bf p}) e^{ipx}+
\right. \nonumber\\
&&\left. \qquad\qquad+\theta (t_1 -t_2) v^2_{\sigma} ({\bf p})
\otimes \overline  v^2_{\sigma} ({\bf p}) e^{-ipx}\right  ]  +
\mbox{parity-transformed}\,
\sim \delta^{(4)} (x_2 -x_1)\,,\nonumber
\end{eqnarray}
for the bi-vector fields,
see~\cite{dv-hpa} for notation.
The reasons were that the Weinberg equation propagates both causal and tachyonic solutions~[20].} The case of the complex-valued vector field will be reported in a separate publication.


{\it Indefinite metrics.}
Usually, one considers the {\it hermitian} field operator in the pseudo-Euclidean netric for the electromagnetic potential:
\begin{equation}
A_\mu = \sum_{\lambda}^{} \int {d^3 {\bf k}\over (2\pi)^3 2E_k} 
[\epsilon_\mu (k,\lambda) a_\lambda ({\bf k}) +\epsilon_\mu^\ast (k,\lambda)
a_\lambda^\dagger ({\bf k})]
\end{equation}
with {\it all} four polarizations to be independent ones. Next, one introduces the Lorentz condition in the weak form
\begin{equation}
[a_{0_t} ({\bf k}) - a_0 ({\bf k})] \vert \phi> =0
\end{equation} 
and the indefinite metrics in the Fock space~\cite[p.90 of the Russian edition]{Bogol}:
$a_{0_t}^\ast = -a_{0_t}$ and $\eta a_\lambda = -a^\lambda \eta$, $\eta^2 =1$,
in order to get the correct sign in the energy-momentum vector
and to not have the problem with the vacuum average.

We observe:
1) that the indefinite metric problems may appear even on the massive level
in the Stueckelberg formalism; 2) The Stueckelberg theory has a good massless limit for propagators, and it reproduces the handling of the indefinite metric in the massless limit (the electromagnetic 4-potential case); 3) we generalized the Stueckelberg formalism (considering, at least, two equations); instead of charge-conjugate solutions we may consider the $P-$  or $CP-$ conjugates. The potential field becomes to be the complex-valued field, that may justify the introduction of the anti-hermitian amplitudes.

\vspace*{-10mm}

\section{Conclusions}

\begin{itemize}

\item
The $(1/2,1/2)$ representation contains both the  spin-1 and spin-0
states (cf. with the Stueckelberg formalism).

\item
Unless we take into account the fourth state (the ``time-like" state, or
the spin-0 state) the set of 4-vectors is {\it not} a complete set in a mathematical sense.

\item
We cannot remove terms like $(\partial_\mu B^\ast_\mu)(\partial_\nu B_\nu)$ 
terms from the Lagrangian and dynamical invariants unless apply the Fermi 
method, i.~e., manually. The Lorentz condition applies only to the spin 1 states.

\item
We have some additional terms in the expressions of the energy-momentum vector (and, accordingly, of the 4-current and the Pauli-Lunbanski vectors), which are the consequence of the impossibility to apply the Lorentz condition for spin-0 states.

\item
Helicity vectors are not eigenvectors of the parity operator. Meanwhile, the parity is a ``good" quantum number, $[{\cal P}, {\cal H}]_- =0$ in the Fock space.

\item
We are able to describe states of different masses in this representation from the beginning.

\item
Various-type field operators can be constructed in the $(1/2,1/2)$ representation space. For instance, they can contain $C$, $P$ and $CP$ conjugate states.
Even if $b_\lambda^\dagger =a_\lambda^\dagger$ 
we can have complex 4-vector fields.\footnote{Perhaps, there are some relations  to the old Weyl idea, recently employed by Kharkov physicists. The sense of this idea is the unification
through the complex potential.}
We found the relations between creation, annihilation operators for different types of the field operators $B_\mu$.

\item
Propagators have good behavious in the massless limit as opposed to those of the Proca theory.

\end{itemize}

The detailed explanations of several claims presented in this talk will be given in journal publications.

\acknowledgments
I am grateful to Profs. Y. S. Kim, S. I. Kruglov, 
V. Onoochin, Z. Oziewicz, W. Rodrigues, R. Santilli, R. Yamaleev and participants of the recent conferences for
useful discussions. 

\section*{Appendix}

In the Dirac case we should assume the following relation in the field operator:
\begin{equation}
\sum_{\lambda}^{} v_\lambda (k) b_\lambda^\dagger (k) = \sum_{\lambda}^{} u_\lambda (-k) a_\lambda (-k)\,.\label{dcop}
\end{equation}
We know that~\cite{Ryder}
\begin{mathletters}
\begin{eqnarray}
\bar u_\mu (k) u_\lambda (k) &=& +m \delta_{\mu\lambda}\,,\\
\bar u_\mu (k) u_\lambda (-k) &=& 0\,,\\
\bar v_\mu (k) v_\lambda (k) &=& -m \delta_{\mu\lambda}\,,\\
\bar v_\mu (k) u_\lambda (k) &=& 0\,,
\end{eqnarray}
\end{mathletters}
but we need $\Lambda_{\mu\lambda} (k) = \bar v_\mu (k) u_\lambda (-k)$.
By direct calculations,  we find
\begin{equation}
-mb_\mu^\dagger (k) = \sum_{\nu}^{} \Lambda_{\mu\lambda} (k) a_\lambda (-k)\,.
\end{equation}
Hence, $\Lambda_{\mu\lambda} = -im ({\bbox \sigma}\cdot {\bf n})_{\mu\lambda}$
and 
\begin{equation}
b_\mu^\dagger (k) = i({\bbox\sigma}\cdot {\bf n})_{\mu\lambda} a_\lambda (-k)\,.
\end{equation}
Multiplying (\ref{dcop}) by $\bar u_\mu (-k)$ we obtain
\begin{equation}
a_\mu (-k) = -i ({\bbox \sigma} \cdot {\bf n})_{\mu\lambda} b_\lambda^\dagger (k)\,.
\end{equation}
The equations (60) and (61) are self-consistent.

In the $(1,0)\oplus (0,1)$ representation we have somewhat different situation:
\begin{equation}
a_\mu (k) = [1-2({\bf S}\cdot {\bf n})^2]_{\mu\lambda} a_\lambda (-k)\,. 
\end{equation}
This signifies that in order to construct the Sankaranarayanan-Good field operator (which was used by Ahluwalia, Johnson and Goldman [Phys. Lett. B (1993)], it satisfies 
$[\gamma_{\mu\nu} \partial_\mu \partial_\nu - {(i\partial/\partial t)\over E} 
m^2 ] \Psi =0$, we need additional postulates, which are possibly related to 
the recent Santilli discoveries (see, for instance, ref.~\cite{Santilli}).

\end{document}